\documentclass[twoside,11pt]{3ssmmp} 

\usepackage{fancyhdr}    
\usepackage{graphicx}
\usepackage{amsmath,amssymb}
\usepackage{floatflt}

       \newcommand {\G }  {\Gamma}
\newcommand {\dl}   {\delta}       \newcommand {\e }  {\epsilon}
        
\newcommand {\ve}   {\varepsilon}  
\newcommand {\lm}   {\lambda}      \newcommand {\m }  {\mu}
          
\newcommand {\s }   {\sigma}      

         \newcommand {\om}  {\omega}
      \newcommand {\Om}  {\Omega}
\newcommand {\Th}   {\Theta}       
\newcommand {\pl}   {\partial}     
\newcommand   {\diag}{{\sf\,diag\,}}
\newcommand {\MO}  {{\mathbb O}}     \newcommand {\MR}  {{\mathbb R}}
\newcommand {\MS}  {{\mathbb S}}     \newcommand {\MP}  {{\mathbb P}}
\newcommand {\Bb}  {\boldsymbol{b}}
\newcommand {\Go}  {\mathfrak{o}}    \newcommand {\Gs}  {\mathfrak{s}}

\begin{document}
 \baselineskip=11pt

\title{Introduction to the Geometric Theory of Defects\hspace{.25mm}\thanks{\,Work supported by
grants RFBR-02-01-01084 and NSH-00-15-96073}}
\author{\bf{M. O. Katanaev}\hspace{.25mm}\thanks{\,e-mail address: katanaev@mi.ras.ru}
\\ \normalsize{Steklov Mathematical Institute, Gubkin St. 8, Moscow, 119991, Russia} }

\date{17 January 2005}

\maketitle

\begin{abstract}
We describe defects -- dislocations and disclinations -- in the framework
of Rie\-mann--Cartan geometry. Curvature and torsion tensors are interpreted
as surface densities of Frank and Burgers vectors, respectively. Equations of
nonlinear elasticity theory are used to fix the coordinate system. The Lorentz
gauge yields equations for the principal chiral $\MS\MO(3)$-field. In the absence
of defects the geometric model reduces to the elasticity theory for the displacement
vector field and to the principal chiral $\MS\MO(3)$-field model for the spin structure.
\end{abstract}
\section{Introduction}
Many solids have a crystalline structure. However, ideal crystals are
absent in Nature, and most of their physical properties, such as
plasticity, melting, growth, etc., are defined by defects of the crystalline
structure. Therefore a study of defects is the actual scientific problem
important for applications in the first place. A broad experimental and
theoretical investigations of defects in crystals started in the thirties
of the last century and are continued to nowadays. At present a fundamental
theory of defects is absent in spite of the existence of dozens of monographs
and thousands of articles.

One of the most promising approach to the theory of defects is based on
Riemann--Cartan geometry, which is given by nontrivial metric and torsion.
In this approach a crystal is considered as continuous elastic media with
a spin structure. If a displacement vector field is a smooth function then
there are only elastic stresses corresponding to diffeomorphisms of the
Euclidean space. If a displacement vector field have discontinuities then
we are saying that there are defects in the elastic structure. Defects in
the elastic structure are called dislocations and lead to the appearance
of nontrivial geometry. Precisely, they correspond to nonzero torsion tensor
which is equal to the surface density of the Burgers vector.

The idea to relate torsion to dislocations appeared in the fifties \cite{Kondo52}.
This approach is being successfully developed up to now (note reviews \cite{SedBer67})
and called often the gauge theory of dislocations. A similar approach is
developed also in gravity \cite{HeMcMiNe95}. It is interesting to note that
E.~Cartan introduced torsion in geometry \cite{Cartan22} having analogy
with mechanics of elastic media.

The gauge approach to the theory of defects is developed successfully,
and interesting results are obtained in this way \cite{Malysh00}
Let us note in this connection two respects in which the approach proposed
below is essentially different. In the gauge models of dislocations based
on the translational group or on the semidirect product of rotational group
on translations one chooses usually distortion and displacement field as
independent variables. It is always possible to fix the invariance with
respect to local translations in such a way that the displacement field
becomes zero because the displacement field moves by simple translation
under the action of the translational group. In this sense the displacement
field is the gauge parameter of local translations, and physical observables
do not depend on it in the gauge invariant models.

The other disadvantage of the gauge approach is the equations of equilibrium.
One considers usually equations of Einstein type for distortion or
vielbein with the right hand side depending on the stress tensor. This
appears unacceptable from the physical point of view because of the following
reason. Consider, for example, one straight edge dislocation. In this case
the elastic stress field differs from zero everywhere. Then the torsion
tensor (or curvature) is also nontrivial due to the equations of equilibrium.
This is wrong from our point of view. Really, consider a domain of media
outside the cutting surface and look at the process of creation of the edge
dislocation. The chosen domain was the part of the Euclidean space with
identically zero torsion and curvature before the defect creation. It is
clear that torsion and curvature remain zero because the process of
dislocation formation is a diffeomorphism for the considered domain.
Besides, the cutting surface may be chosen arbitrary for the defect formation
leaving the axis of dislocation unchanged. We deduce from this that torsion
and curvature must be zero everywhere except the axis of dislocation.
In other words the elasticity stress tensor can not be the source of
dislocations. To avoid the appearing contradiction we propose a cardinal
way out: we do not use the displacement field as an independent variable
at all. It does not mean that the displacement field does not exist in real
crystals. In the proposed approach the displacement field exists in those
regions of media which do not contain cores of dislocations, and it can be
computed. In this case it satisfies the equations of nonlinear elasticity
theory.

The proposed geometric approach allows one to include into consideration
other defects which do not relate directly to defects in elastic structure.
The intensive investigations of these defects were conducted in parallel
with the study of dislocations. The point is that many solids do not only
have elastic properties but possess a spin structure. For example, there
are ferromagnets, liquid crystals, spin glasses, etc.
In this case there are defects in the spin structure which are called
disclinations \cite{Frank58}. They arise when the director field has
discontinuities. The presence of disclinations is also connected to nontrivial
geometry. Namely, the curvature tensor equals to the surface density of the
Frank vector. The gauge approach based on the rotational group $\MS\MO(3)$
was also used for describing disclinations \cite{DzyVol78}.
$\MS\MO(3)$-gauge models of spin glasses with defects were considered in
\cite{Hertz78}.

The geometric theory of static distribution of defects which describes both
types of defects -- dislocation and disclinations -- from a single point of view
was proposed in \cite{KatVol92}. In contrast to other approaches we have
vielbein and $\MS\MO(3)$-connection as the only independent variables.
Torsion and curvature tensors have direct physical meaning as the
surface densities of dislocations and disclinations, respectively.
Covariant equations of equilibrium for vielbein and $\MS\MO(3)$-connection
similar to those in a gravity model with torsion are postulated. To define
the solution uniquely we must fix the coordinate system (fix the gauge)
because any solution of the equations of equilibrium is defined up to
general coordinate transformations and local $\MS\MO(3)$-rotations.
The elastic gauge for the vielbein \cite{Katana03} and Lorentz gauge for the
$\MS\MO(3)$-connection \cite{Katana04} were proposed recently. We stress
that the notions of displacement vector and rotational angle are absent in our
approach at all. These notions can be introduced only in those domains where
defects are absent. In this case equations for vielbein and
$\MS\MO(3)$-connection are satisfied identically, the elastic gauge reduces
to the equations of nonlinear elasticity theory for the displacement vector,
and the Lorentz gauge yields equations for the principal chiral
$\MS\MO(3)$-field. In other words, to fix the coordinate system we choose two
fundamental models: elasticity theory and the principal chiral field model.
\section{Elastic deformations                          \label{seldef}}
We consider infinite three dimensional elastic media. Suppose that
undeformed media in the absence of defects is invariant under translations
and rotations in some coordinate system. Then the media in this coordinate
system $y^i$, $i=1,2,3$, is described by the Euclidean metric
$\dl_{ij}={\diag}(+++)$, and the system of coordinates is called Cartesian.
Thus in the undeformed state we have the Euclidean space $\MR^3$ with a given
Cartesian coordinate system. We assume also that torsion
in the media equals zero.

Let a point of media has coordinates $y^i$ in the ground state. After
deformation this point will have coordinates, see Fig.~\ref{feldef},
\begin{equation}                                        \label{eeldef}
  y^i\rightarrow x^i(y)=y^i+u^i(x)
\end{equation}
in the initial coordinate system. The inverse notations are used in the
elasticity theory. One writes usually $x^i\rightarrow y^i=x^i+u^i(x)$.
These are equivalent writings because both coordinate systems $x^i$ and $y^i$
cover the whole $\MR^3$. However in the theory of defects considered in the
next sections the situation is different. In a general case the elastic media
fills the whole Euclidean space only in the final state. Here and in what
follows we assume that fields depend on coordinates $x$ which are coordinates
of points of media after the deformation and cover the whole Euclidean space
$\MR^3$ by assumption. In the presence of dislocations the coordinates
$y^i$ do not cover the whole $\MR^3$ in a general case because part of the
media may be removed or, inversely, added. Therefore the system of coordinates
related to points of the media after an elastic deformation and defect creation
is more preferable.
\begin{floatingfigure}{.45\textwidth}
\includegraphics[width=.4\textwidth]{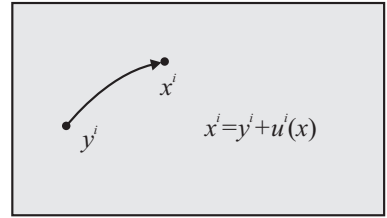}
 \caption{Elastic deformations}
 \label{feldef}
\end{floatingfigure}

In the linear elasticity theory relative deformations are assumed to be small
$\pl_j u^i\ll1$. Then the functions $u^i(x)=u^i(y(x))$ are components of
a vector field which is called the displacement vector field and is the
basic variable in the elasticity theory.

In the absence of defects we assume that the displacement field is a smooth
vector field in the Euclidean space $\MR^3$. The presence of discontinuities
and singularities in displacement field is interpreted as the presence of
defects.

We shall consider only static deformations in what follows when displacement
field $u^i$ does not depend on time. Then the basic equations of equilibrium
for small deformations are (see, for example, \cite{LanLif70})
\begin{align}                                        \label{estati}
  \pl_j\s^{ji}+f^i&=0,
\\                                                      \label{eHook}
  \s^{ij}&=\lm\dl^{ij}\e_k{}^k+2\m\e^{ij},
\end{align}
where $\s^{ji}$ is the stress tensor which is assumed to be symmetric.
The tensor of small deformations $\e_{ij}$ is given by the symmetrized
partial derivative of the displacement vector
\begin{equation}                                        \label{edefte}
  \e_{ij}=\frac12(\pl_i u_j+\pl_j u_i).
\end{equation}
Lowering and raising of the Latin indices is performed with the Euclidean
metric $\dl_{ij}$ and its inverse $\dl^{ij}$.
The letters $\lm$ and $\mu$ denote constants characterizing elastic properties
of media and are called Lame coefficients. Functions $f^i(x)$ describe
total density of nonelastic forces inside the media. We assume that such
forces are absent in what follows, $f^j(x)=0$. Equation (\ref{estati}) is
Newton's law, and Eq.~(\ref{eHook}) is Hook's law relating stresses
with deformations.

In Cartesian coordinate system for small deformations the difference between
upper and lower indices disappears because raising and lowering of indices
is performed with the help of the Euclidean metric. One usually forgets about
this difference due to this reason, and this is fully justified. But in the
presence of defects the notion of Cartesian coordinate system and Euclidean
metric is absent, and raising and lowering of indices are performed with the
help of Riemannian metric. Therefore we distinguish upper and lower indices
as it is accepted in differential geometry having in mind the following
transition to elastic media with defects.

The main problem in the linear elasticity theory is the solution of the
second order equations for displacement vector which arise after
substitution of (\ref{eHook}) into (\ref{estati}) with some boundary
conditions. Many known solutions are in good agreement with experiment.
Therefore one may say that equations (\ref{eHook}), (\ref{estati}) have
a solid experimental background.

Let us look at the elastic deformations from the point of view of differential
geometry (see, for example, \cite{DuNoFo98E}). From mathematical standpoint the
map (\ref{eeldef}) by itself is
the diffeomorphism of the Euclidean space $\MR^3$. In this case the Euclidean
metric $\dl_{ij}$ is induced by the map $y^i\rightarrow x^i$. It means that
in the deformed state  the metric in the linear approximation has the form
\begin{equation}                                        \label{emetri}
  g_{ij}(x)=\frac{\pl y^k}{\pl x^i}\frac{\pl y^l}{\pl x^j}
     \dl_{kl}\approx\dl_{ij}-\pl_iu_j-\pl_ju_i=\dl_{ij}-2\e_{ij},
\end{equation}
i.e.\ it is defined by the tensor of small deformations (\ref{edefte}).
Note that in the linear approximation $\e_{ij}(x)=\e_{ij}(y)$ and
$\pl u_j/\pl x^i=\pl u_j/\pl y^i$.

In Riemann geometry the metric defines uniquely the Levi--Civita connection
$\widetilde\G_{ij}{}^k(x)$ (Christoffel's symbols)
\begin{equation}                                        \label{etigam}
  \widetilde\G_{ijk}=\frac12(\pl_i g_{jk}+\pl_j g_{ik}-\pl_k g_{ij}).
\end{equation}
We can compute the curvature tensor
\begin{equation}                                        \label{ecurva}
  \widetilde R_{ijk}{}^l=\pl_i\widetilde\G_{jk}{}^l-
  \widetilde\G_{ik}{}^m\widetilde\G_{jm}{}^l-(i\leftrightarrow j),
\end{equation}
for these symbols. This tensor equals identically zero, $\widetilde R_{ijk}{}^l(x)=0$,
because curvature of the Euclidean space is zero, and the map $y^i\rightarrow x^i$ is a
diffeomorphism. The torsion tensor equals zero for the same reason.
Thus an elastic deformation of media corresponds to trivial Riemann--Cartan
geometry because curvature and torsion tensors are equal to zero.

The physical interpretation of the metric (\ref{emetri}) is the following.
External observer fixes Cartesian coordinate system corresponding to the
ground undeformed state of media. The media is deformed afterwards, and
external observer discovers that the metric in this coordinate system
becomes nontrivial. If we assume that elastic perturbations in media
(phonons) propagate along extremals $x^k(t)$ (lines of minimal length), then in
the deformed media their trajectories will be defined by equations
\begin{equation*}                                     \label{exteqr}
  \ddot x^k=-\widetilde\G_{ij}{}^k\dot x^i\dot x^j,
\end{equation*}
where dots denote differentiation with respect to a canonical parameter $t$.
Trajectories of phonons will be not now straight lines
because Christoffel's symbols are nontrivial, $\widetilde\G_{ij}{}^k\ne0$.
In this sense the metric (\ref{emetri}) is observable. Here we see the
essential role of the Cartesian coordinate system $y^i$ defined by the
undeformed state with which the measurement process is connected.

Assume that the metric $g_{ij}(x)$ given in the Cartesian coordinates
corresponds to some state of elastic media without defects. In this
case the displacement vector is defined by the system of equations
(\ref{emetri}), and its integrability conditions are the equality of the
curvature tensor to zero. In the linear approximation these conditions
are known in the elasticity theory as the Saint--Venant integrability
conditions.

Let us make the remark important for the following consideration.
For appropriate boundary conditions the solution of the elasticity theory
equations (\ref{estati}), (\ref{eHook}) is unique. From the geometric
viewpoint it means that elasticity theory fixes diffeomorphisms. This
fact will be used in the geometric theory of defects. Equations of
nonlinear elasticity theory written in terms of metric or vielbein
will be used for fixing the coordinate system.
\section{Dislocations                                  \label{sdislo}}
\begin{figure}[h,t]
\hfill\includegraphics[width=.35\textwidth]{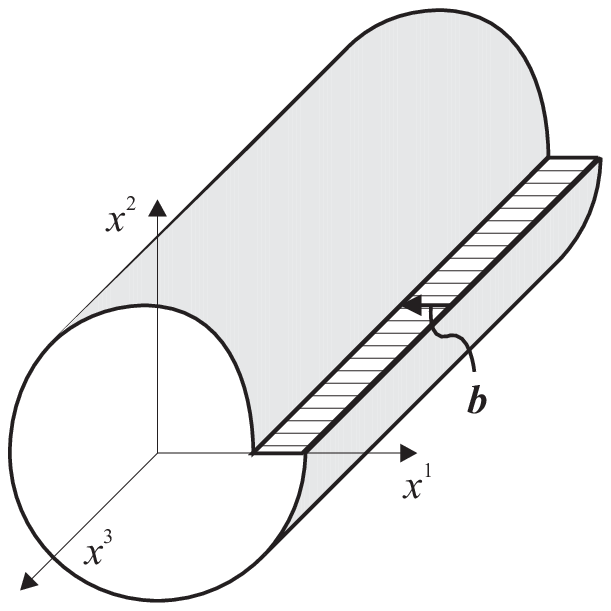}
\hfill \hspace*{.06\textwidth} \hfill
\includegraphics[width=.35\textwidth]{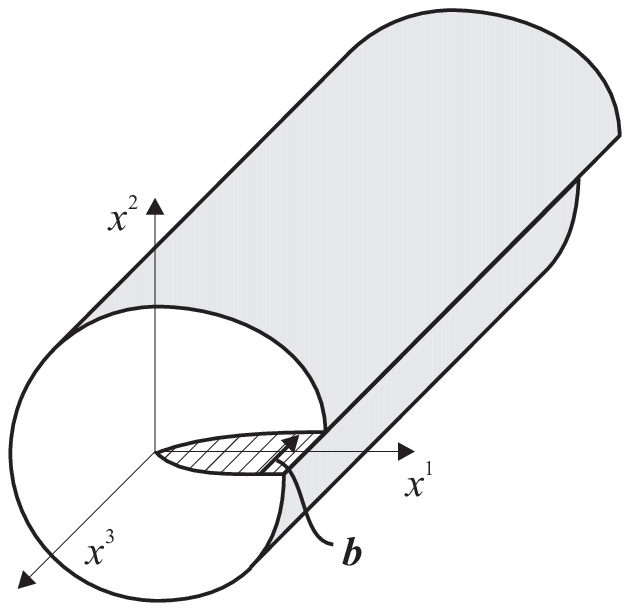}
\hfill {}
\\
\parbox[t]{.47\textwidth}{(\textit{a}) The edge dislocation. The Burgers
           vector $\Bb$ is perpendicular to the dislocation line.}
\hfill
\parbox[t]{.47\textwidth}{(\textit{b}) The screw dislocation. The Burgers
           vector $\Bb$ is parallel to the dislocation line.}
\\
\centering \caption{\label{fdislo} Straight linear dislocations.}
\end{figure}

We start with description of linear dislocations in elastic media
(see, for example, \cite{LanLif70,Kosevi81}). The simplest and widely
spreaded examples of linear dislocations are shown in Fig.~\ref{fdislo}.
Cut the media along the half plane $x^2=0$, $x^1>0$. Move the upper part
of the media located over the cut $x^2>0$, $x^1>0$ on the vector $\Bb$
towards the dislocation axis $x^3$, and glue the cutting surfaces.
The vector $\Bb$ is called the Burgers vector. In a general case the
Burgers vector may not be constant on the cut. For the edge dislocation
it varies from zero to some constant value $\Bb$ as it moves from the
dislocation axis. After the gluing the media comes to the equilibrium state
which is called the edge dislocation shown in Fig.~\ref{fdislo},\textit{a}.
If the Burgers vector is parallel to the dislocation line then it is called
the screw dislocation, Fig.~\ref{fdislo},\textit{b}.

One and the same dislocation can be made in different ways. For example,
if the Burgers vector is perpendicular to the cutting plane and directed
from it in the considered cases then the produced cavity must be filled with
media before the gluing. One can easily imagine that the edge dislocation
is also obtained as the result but rotated by the angle $\pi/2$ around the
axis $x^3$. This example shows that a dislocation is characterized not
by the cutting surface but by the dislocation line and the Burgers vector.

From topological point of view the media containing several dislocations
or even the infinite number of them represents itself the Euclidean space
$\MR^3$. In contrast to elastic deformations the displacement vector
in the presence of dislocations is no longer a smooth function because
of the presence of cutting surfaces. At the same time we assume that
partial derivatives of the displacement vector $\pl_j u^i$ (the distortion
tensor) are smooth functions on the cutting surface. This assumption
is justified physically because these derivatives define the deformation
tensor (\ref{edefte}). In its turn partial derivatives of deformation tensor
must exist and be smooth functions in the equilibrium state everywhere
except, possibly, the axis of dislocation because otherwise the equations
of equilibrium (\ref{estati}) do not have meaning. We assume that the metric
and vielbein are smooth functions everywhere in $\MR^3$ except, may be,
dislocation axes because the deformation tensor defines the induced metric
(\ref{emetri}).

The main idea of the geometric approach reduces to the following. To describe
single dislocations in the framework of elasticity theory we must solve
equations for the displacement vector with some boundary conditions on the
cuts. For small number of dislocations this is possible. However with
increasing number of dislocations the boundary conditions become so
complicated that the solution of the problem becomes unreal. Besides,
one and the same dislocation may be produced by different cuts which
lead to ambiguity in the displacement vector field. Another shortcoming
of this approach is that it can not be applied for description of
continuous distribution of dislocations because in this case the displacement
vector field does not exist at all for the reason that it must have
discontinuities at every point. In the geometric approach the basic
variable is the vielbein which is a smooth function everywhere except,
possibly, dislocation axes by assumption. We postulate new equations
for the vielbein (see \cite{KatVol92}). In the geometric approach the
transition from finite number of dislocations to their continuous
distribution is simple and natural. In that way the smoothing of
singularities occur on dislocation axes in analogy with smoothing of mass
distribution for point particles when going to continuous media.

Let us start to build the formalism of the geometric approach. In a general
case in the presence of defects we do not have a preferred Cartesian
coordinates frame in the equilibrium state because there is no symmetry.
Therefore we consider arbitrary coordinates $x^\mu$, $\mu=1,2,3$, in
$\MR^3$. Now we are using Greek letters to enumerate coordinates admitting
arbitrary coordinate changes. Then the Burgers vector can be expressed
as the integral of the displacement vector
\begin{equation}                                        \label{eBurge}
  \oint_Cdx^\mu\pl_\mu u^i(x)=-\oint_Cdx^\mu\pl_\mu y^i(x)=-b^i,
\end{equation}
where $C$ is a closed contour surrounding the dislocation axis, Fig.~\ref{fburco}.
\begin{floatingfigure}{.45\textwidth}
\includegraphics[width=.4\textwidth]{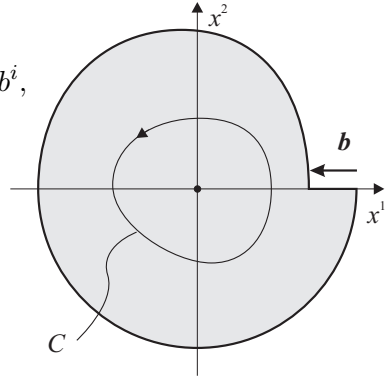}
 \caption{The section of the media with the edge dislocation. $C$ is the
 integration contour for the Burgers vector $\Bb$.}
 \label{fburco}
\end{floatingfigure}
This integral is invariant with respect to arbitrary coordinate
transformations $x^\mu\rightarrow x^{\mu'}(x)$ and covariant under global
$\MS\MO(3)$-rotations of $y^i$. Here components of the displacement vector
field  $u^i(x)$ are considered with respect to the orthonormal basis in
the tangent space $u=u^i e_i$. If components of the displacement vector
field were considered with respect to the coordinate basis $u=u^\mu\pl_\mu$,
then the invariance of the integral (\ref{eBurge}) under general coordinate
changes is violated.

In the geometric approach we introduce new independent variable -- the vielbein
-- instead of partial derivatives $\pl_\mu u^i$
\begin{equation}                                        \label{edevid}
  e_\mu{}^i(x)=\begin{cases} \pl_\mu y^i, &\text{outside the cut,}\\
               \lim\pl_\mu y^i, &\text{on the cut.}\end{cases}
\end{equation}
The vielbein is a smooth function on the cut by construction. Note that
if the vielbein was simply defined as partial derivative $\pl_\mu y^i$,
then it would have the $\dl$-function singularity on the cut because functions
$y^i(x)$ have a jump. Then the Burgers vector can be expressed through the
integral over a surface $S$ having contour $C$ as the boundary

\begin{equation}                                        \label{eBurg2}
  \oint_Cdx^\mu e_\mu{}^i=\int\!\!\int_Sdx^\mu\wedge dx^\nu
  (\pl_\mu e_\nu{}^i-\pl_\nu e_\mu{}^i)=b^i,
\end{equation}
where $dx^\mu\wedge dx^\nu$ is the surface element. From the definition of
the vielbein (\ref{edevid}) we see that the integrand equals zero everywhere
except the axis of dislocation. For the edge dislocation with constant
Burgers vector the integrand has $\dl$-function singularity at the origin.
The criterion for the presence of dislocation is a violation of integrability
conditions for the system of equations $\pl_\mu y^i=e_\mu{}^i$:
\begin{equation}                                        \label{eintco}
  \pl_\mu e_\nu{}^i-\pl_\nu e_\mu{}^i\ne0.
\end{equation}
If dislocations are absent then functions $y^i(x)$ exist and define
transformation to the Cartesian coordinates frame.

In the geometric theory of defects the field $e_\mu{}^i$ is identified
with the vielbein. Next, compare the integrand in (\ref{eBurg2}) with the
expression for torsion in Cartan variables
\begin{equation}                                           \label{ecurcv}
  T_{\mu\nu}{}^i=\pl_\mu e_\nu{}^i-e_\mu{}^j\om_{\nu j}{}^i
                    -(\mu\leftrightarrow\nu),
\end{equation}
They differ only by terms containing $\MS\MO(3)$-connection $\om_{\nu j}{}^i(x)$.
This is the ground for the introduction of the following postulate. In the
geometric theory of defects the Burgers vector corresponding to a surface $S$
is defined by the integral of the torsion tensor
\begin{equation*}
  b^i=\int\!\!\int_S dx^\mu\wedge dx^\nu T_{\mu\nu}{}^i.
\end{equation*}
This definition is invariant with respect to general coordinate transformations
of $x^\mu$ and covariant with respect to global rotations. Thus the torsion
tensor has straightforward physical interpretation: it equals the surface
density of the Burgers vector.

Physical interpretation of the $\MS\MO(3)$-connection will be given in the
next section, and now we show how this definition reduces to the expression
for the Burgers vector (\ref{eBurg2}) obtained within the elasticity theory.
If the curvature tensor for $\MS\MO(3)$-connection equals zero, then the
connection is locally trivial, and there exists such $\MS\MO(3)$ rotation
that $\om_{\mu i}{}^j=0$. In that case we return to expression (\ref{eBurg2}).

If $\MS\MO(3)$-connection is zero and vielbein is a smooth function
then the Burgers vector corresponds uniquely to every contour. In this case
it can be expressed as the surface integral of the torsion tensor. The
surface integral depends only on the boundary contour but not on the
surface due to the Stokes theorem.
\begin{floatingfigure}{.45\textwidth}
\includegraphics[width=.4\textwidth]{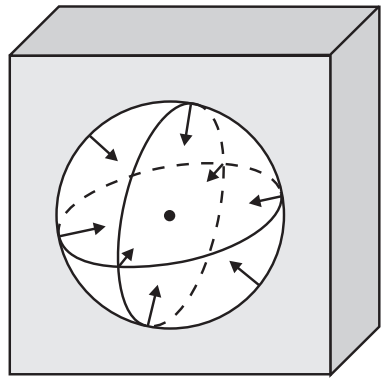}
 \caption{Point defect -- vacancy -- appears when a ball is cut out from the
 media, and the boundary sphere is shrank to a point.}
 \label{fpoide}
\end{floatingfigure}

We have shown that the presence of linear defects results in a nontrivial
torsion tensor. In the geometric theory of defects the equality of torsion
tensor to zero $T_{\mu\nu}{}^i=0$ is naturally considered as the criterion
for the absence of dislocations. Then under the name dislocation fall not
only linear dislocations but, in fact,
arbitrary defects in elastic media. For example, point defects: vacancies
and impurities are also dislocations. In the first case we cut out a ball
from the Euclidean space $\MR^3$ and then shrink the boundary sphere to a
point, Fig.~\ref{fpoide}. In the case of impurity a point of the Euclidean
space is blown up to a sphere and the produced cavity is filled with the media.
Point defects are characterized by the mass of the removed or added media
which is also defined by the vielbein \cite{KatVol92}
\begin{equation}                                        \label{emass}
  M=\rho_0\int\!\!\int\!\!\int_{\MR^3} d^3x\left(\det e_\mu{}^i
  -\det\overset\circ e_\mu{}^i\right),
  ~~~~\overset\circ e_\mu{}^i=\pl_\mu y^i,
\end{equation}
where $y^i(x)$ are the transition functions to Cartesian coordinate frame
in $\MR^3$, and $\rho_0$ is the density of the media which is supposed to
be constant. The mass is defined by the difference of two integrals each of
them being divergent separately. The first integral equals to the volume
of the media with defects and the second is equal to the volume of the
Euclidean space. According to the given definition the mass of an impurity
is positive because the matter is added to the media, and the mass of a
vacancy is negative since part of the media is removed.
The torsion tensor for a vacancy or impurity equals zero
everywhere except one point where it has a $\dl$-function singularity.
For point defects the notion of the Burgers vector is absent.

In three dimensional space surface defects may also exist along with point
and line dislocations. In the geometric approach all of them are called
dislocations because they correspond to nontrivial torsion.
\section{Disclinations                                 \label{sdiscl}}
We relate dislocations to nontrivial torsion tensor in the preceeding
section. To this end we introduced the $\MS\MO(3)$-connection.
In the present section we show that the curvature tensor for the
$\MS\MO(3)$-connection defines the surface density of the Frank vector
characterizing other well known defects -- disclinations in the spin
structure \cite{LanLif70}.

\begin{figure}[h,t]
\hfill\includegraphics[width=.35\textwidth]{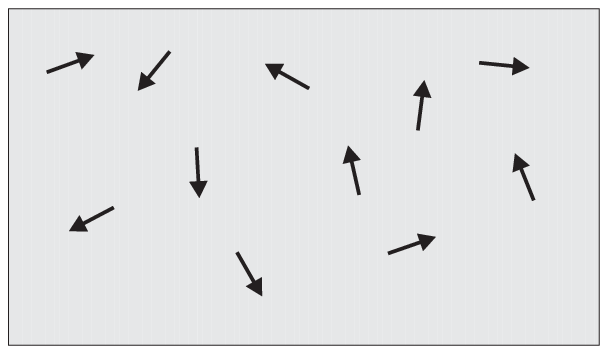}
\hfill  \hspace*{.06\textwidth} \hfill
\includegraphics[width=.35\textwidth]{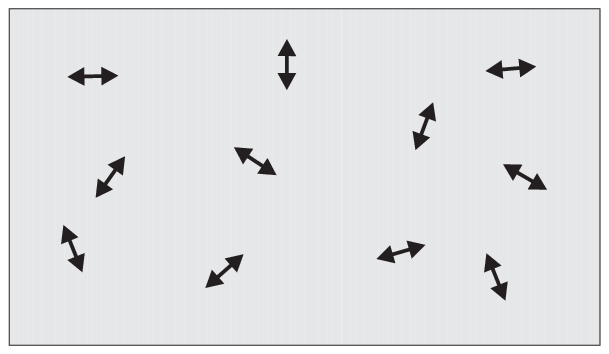}
\hfill {} \\
\parbox[t]{.47\textwidth}{\centering (\textit{a}) Ferromagnet}
\hfill
\parbox[t]{.47\textwidth}{\centering (\textit{b}) Liquid crystal}
\\
\centering \caption{\label{fspstr} Examples of media with the spin structure}
\end{figure}

Let the unit vector field $n^i(x)$, $(n^in_i=1)$, be given in all
points of media. For example, $n^i$ has the meaning of magnetic moment
located at each point of the media for ferromagnets,
Fig.~\ref{fspstr},\textit{a}. For nematic liquid crystals the unit vector
field $n^i$ with the equivalence relation $n^i\sim-n^i$ describes
the director field, Fig.~\ref{fspstr}, \textit{b}.

Let us fix some direction in the media $n_0^i$. Then the field $n^i(x)$
at a point $x$ can be uniquely defined by the field $\om^{ij}(x)=-\om^{ji}(x)$
taking values in the rotation algebra $\Gs\Go(3)$ (the angle of rotation):
$$
  n^i=n_0^j S_j{}^i(\om),
$$
where $S_j{}^i\in\MS\MO(3)$ is the rotation matrix corresponding to the
algebra element $\om^{ij}$. Here we use the following parameterization of
the rotation group $\MS\MO(3)$ by elements of its algebra (see, for
example, \cite{Katana04})
\begin{equation}                                        \label{elsogr}
  S_i{}^j=(e^{(\om\ve)})_i{}^j=\cos\om\,\dl_i^j
  +\frac{(\om\ve)_i{}^j}\om\sin\om
  +\frac{\om_i\om^j}{\om^2}(1-\cos\om)~          \in\MS\MO(3),
\end{equation}
where $(\om\ve)_i{}^j=\om^k\ve_{ki}{}^j$ and $\om=\sqrt{\om^i\om_i}$ is
the modulus of the vector $\om^i$. The pseudovector
$\om^k=\frac12\om_{ij}\ve^{ijk}$, where $\ve^{ijk}$ is the totally
antisymmetric third rank tensor, $\ve^{123}=1$, is directed along the
rotational axis and its length equals to the angle of rotation. We shall
call the field $\om^{ij}$ a spin structure of the media.

\begin{figure}[h,t]
\hfill\includegraphics[width=.35\textwidth]{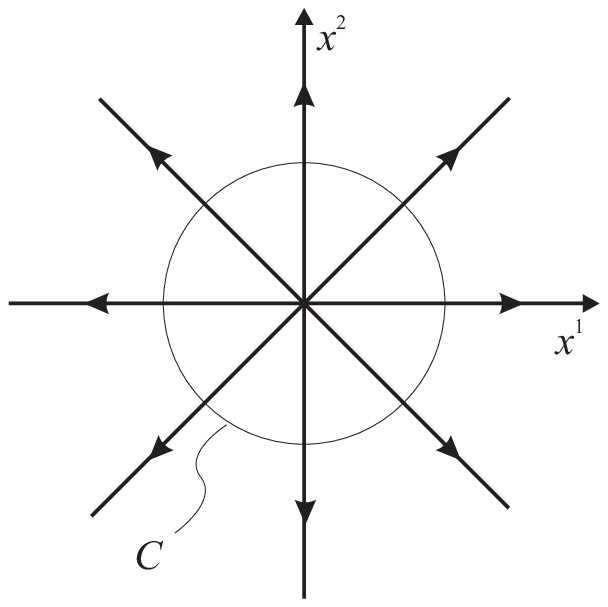}
\hfill  \hspace*{.06\textwidth} \hfill
\includegraphics[width=.35\textwidth]{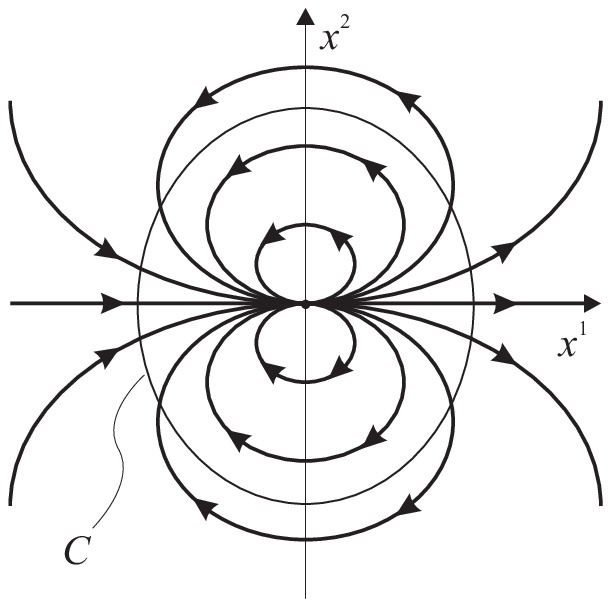}
\hfill {} \\
\parbox[t]{.47\textwidth}{\centering (\textit{a}) $|\Theta|=2\pi$}
\hfill
\parbox[t]{.47\textwidth}{\centering (\textit{b}) $|\Theta|=4\pi$}
\\
\centering \caption{\label{fdiscl} The vector field distributions in the
    plane $x^1,x^2$ for the linear dislocations parallel to the $x^3$ axis.}
\end{figure}

If the media possesses a spin structure then it may have defects called
disclinations. For linear disclinations parallel to the $x^3$ axis
the vector field $n$ lies in the perpendicular plain $x^1,x^2$.
The simplest examples of linear disclinations are shown in Fig.~\ref{fdiscl}.
Every linear disclination is characterized by the Frank vector
\begin{equation}                                        \label{etheta}
  \Th_i=\e_{ijk}\Om^{jk},
\end{equation}
where
\begin{equation}                                        \label{eomega}
  \Om^{ij}=\oint_Cdx^\mu\pl_\mu\om^{ij},
\end{equation}
and the integral is taken along closed contour $C$ surrounding the disclination
axis. The length of the Frank vector is equal to the total angle of
rotation of the field $n^i$ as we go around the disclination.

The vector field $n^i$ defines a map of the Euclidean space into sphere
$n:~\MR^3\rightarrow\MS^2$. For linear disclinations parallel to the
$x^3$ axis this map is restricted to a map of the plane $\MR^2$ into
a circle $\MS^1$. In this case the total angle of rotation
must be obviously a multiple of $2\pi$.

\begin{figure}[h,t]
\hfill\includegraphics[width=.35\textwidth]{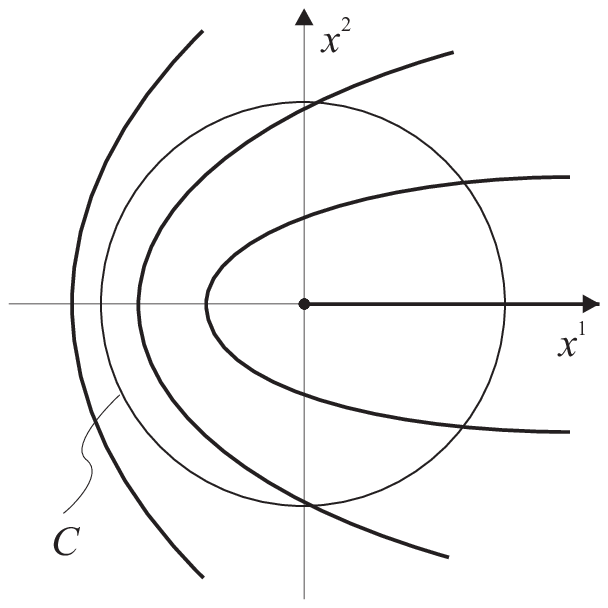}
\hfill  \hspace*{.06\textwidth} \hfill
\includegraphics[width=.35\textwidth]{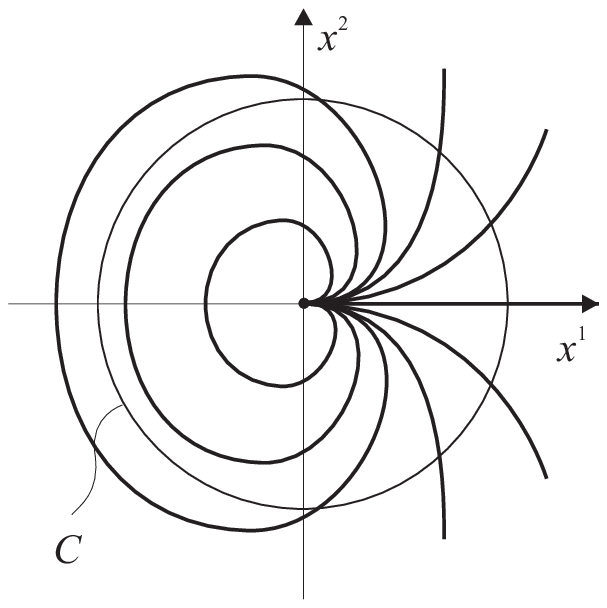}
\hfill {} \\
\parbox[t]{.47\textwidth}{\centering (\textit{a}) $|\Theta|=\pi$}
\hfill
\parbox[t]{.47\textwidth}{\centering (\textit{b}) $|\Theta|=3\pi$}
\\
\centering \caption{\label{fdisc2} The director field distribution
    in the $x^1,x^2$ plane for the linear disclinations parallel to
    the $x^3$ axis.}
\end{figure}

For nematic liquid crystals we have the equivalence relation $n^i\sim-n^i$.
Therefore for linear disclinations parallel to the $x^3$ axis the director
field defines a map of the plane into the projective line,
$n:~\MR^2\rightarrow\MR\MP^1$. In this case the length of the Frank vector
must be a multiple of $\pi$. The corresponding examples of disclinations
are shown in Fig.~\ref{fdisc2}.

As in the case of the displacement field, the field $\om^{ij}(x)$ taking
values in the algebra $\Gs\Go(3)$ is not a smooth function in $\MR^3$ in
the presence of disclinations. Let us make a cut in $\MR^3$ bounded by
the disclination axis. Then the field $\om^{ij}(x)$ may be considered as
smooth in the whole space except the cut. We assume that all partial
derivatives of $\om^{ij}(x)$ have the same limit as far as it approaches
the cut from both sides. Then we define a new field
\begin{equation}                                        \label{edesoc}
  \om_\mu{}^{ij}=\begin{cases}\pl_\mu\om^{ij}, &\text{outside the cut,}\\
                 \lim\pl_\mu\om^{ij}, &\text{on the cut.} \end{cases}
\end{equation}
The functions $\om_\mu{}^{ij}$ are smooth  by construction everywhere except,
may be, the disclination axis. Then the Frank vector may be
given by the surface integral
\begin{equation}                                        \label{eFrank}
  \Om^{ij}=\oint_Cdx^\mu\om_\mu{}^{ij}=
  \int\!\!\int_Sdx^\mu\wedge dx^\nu(\pl_\mu\om_\nu{}^{ij}-\pl_\nu\om_\mu{}^{ij}),
\end{equation}
where $S$ is an arbitrary surface having contour $C$ as the boundary. If the field
$\om_\mu{}^{ij}$ is given then the integrability conditions for the system
of equations $\pl_\mu\om^{ij}=\om_\mu{}^{ij}$ are the equalities
\begin{equation}                                        \label{eomcon}
    \pl_\mu\om_\nu{}^{ij}-\pl_\nu\om_\mu{}^{ij}=0.
\end{equation}
This noncovariant equality yields the criterion for the absence of
disclinations.

In the geometric theory of defects we identify the field $\om_\mu{}^{ij}$
with the $\MS\MO(3)$-connection. In the expression for the curvature in
Cartan variables
\begin{equation}                                             \label{ecucav}
  R_{\mu\nu j}{}^i=\pl_\mu \om_{\nu j}{}^i-\om_{\mu j}{}^k
                       \om_{\nu k}{}^i-(\mu\leftrightarrow\nu),
\end{equation}
the first two terms coincide with (\ref{eomcon}), therefore
we postulate the covariant criterion of the absence of disclinations as
the equality of curvature tensor for $\MS\MO(3)$-connection to zero
\begin{equation*}
    R_{\mu\nu}{}^{ij}=0.
\end{equation*}
Simultaneously, we give the physical interpretation of the curvature tensor
as the surface density of the Frank vector
\begin{equation}                                        \label{eOmega}
    \Om^{ij}=\int\!\!\int dx^\mu\wedge dx^\nu R_{\mu\nu}{}^{ij}.
\end{equation}
This definition reduces to the previous expression of the Frank vector
(\ref{eFrank}) in the case when rotation of vector $n$ takes place in
a fixed plane. In this case rotations are restricted by the subgroup
$\MS\MO(2)\subset\MS\MO(3)$. The quadratic terms in the expression for
the curvature (\ref{ecucav}) disappear because the rotation group $\MS\MO(2)$
is Abelian, and we obtain the previous expression for the Frank vector
(\ref{eFrank}).

Thus we described the media with dislocations (defects of elastic media)
and disclinations (defects in the spin structure) in the framework of
Riemann--Cartan geometry. Here we identified torsion tensor with the
surface density of dislocations and curvature tensor with the surface density
of disclinations. The relations between physical and geometrical notions
are summarized in the Table~\ref{tdegeo}.
\begin{table}[h]
\begin{center}
    \begin{tabular}{|l|l|l|}                                  \hline
    Elastic deformations &$R_{\mu\nu}{}^{ij}=0$&$T_{\mu\nu}{}^i=0$\\ \hline
    Dislocations &$R_{\mu\nu}{}^{ij}=0$&$T_{\mu\nu}{}^i\ne0$\\ \hline
    Disclinations &$R_{\mu\nu}{}^{ij}\ne0$&$T_{\mu\nu}{}^i=0$\\ \hline
    Dislocations and disclinations&$R_{\mu\nu}{}^{ij}\ne0$&$T_{\mu\nu}{}^i\ne0$\\
    \hline
    \end{tabular}
    \caption{\label{tdegeo} The relation between physical and geometrical
    notions in the geometric theory of defects.}
\end{center}
\end{table}
\section{Conclusion}
The geometric theory of defects describes defects in elastic media
(dislocations) and defects in the spin structure (disclinations) from
the unique point of view. This model can be used for description of
single defects as well as their continuous distribution. The geometric
theory of defects is based on the Riemann-Cartan geometry. By definition
torsion and curvature tensors are equal to surface densities of Burgers
and Frank vectors, respectively.

We gave here only the relation between physical and geometric notions.
At the moment the static geometric theory of defects is developed much further.
In \cite{Katana04B} we considered the example of the wedge dislocation both
in the frameworks of ordinary elasticity theory and geometric theory of defects
and compared the results. This example shows that the elasticity theory
reproduces only the linear approximation of the geometric theory of defects.
In contrast to the induced metric obtained within the exact solution
of the linear elasticity theory, the expression for the metric obtained as
the exact solution of the Einstein equations is simpler, defined on the
whole space and for all deficit angles. The obtained expression for the
metric can be checked experimentally.

In \cite{Katana04B} we showed also that the equations of asymmetric elasticity
theory for the Cosserat media are naturally embedded in the geometric theory
of defects as the gauge conditions. We also considered there two problems as an
application of the geometric theory of defects. The first is the scattering of
phonons on a wedge dislocation \cite{Moraes96}. In the eikonal approximation
the problem is reduced to the analysis of extremals for the metric describing
a given dislocation. Equations for extremals are integrated explicitly, and
the scattering angle is found. The second of the considered problems is the
construction of wave functions and energetic spectrum of impurity in the
presence of a wedge dislocation. To this purpose we solved the Schr\"odinger
equation. This problem is mathematically equivalent to solution of the
Schr\"odinger equation for bounded states in the Aharonov--Bohm effect
\cite{Skarzh81}. The explicit dependence of the spectrum from the deficit
angle and elastic properties of media is found.

Equations defining the static distribution of defects are covariant and
have the same form as equations of gravity models with dynamical torsion.
To choose a solution uniquely, one must fix the coordinate system.
To this end the elastic gauge for the vielbein and the Lorentz gauge for
the $\MS\MO(3)$-connection are proposed. If defects are absent then we can
introduce the displacement vector field and the field of the spin structure.
In this case equations of equilibrium are identically satisfied, and the
gauge conditions reduce to the equations of the elasticity theory and of
the principal chiral $\MS\MO(3)$-field. In this way the geometric theory
of defects incorporate the elasticity theory and the model of principal
chiral field.

In a definite sense the elastic gauge represents the equations of
nonlinear elasticity theory. Nonlinearity is introduced in elasticity theory
in two ways. First, the deformation tensor is defined through the induced
metric $\e_{ij}=\frac12(\dl_{ij}-g_{ij})$ instead of its definition by
the linear relation (\ref{edefte}). Then the stress tensor is given by the
infinite series in the displacement vector. Second, Hook's law can be
modified assuming nonlinear dependence of the stress tensor on the
deformation tensor. Hence the elastic gauge is the equations of nonlinear
elasticity theory where the deformation tensor is assumed to be defined
through the induced metric, and Hook's law is kept linear. A generalization
to a more general case when the relation between deformation and stress
tensors becomes nonlinear is obvious.

The geometric theory of static distribution of defects can be also
constructed for the membranes, i.e.\ on a plane $\MR^2$. To this end
one has to consider the Euclidean version \cite{Katana97} of two-dimensional
gravity with torsion \cite{VolKat86}. This model is favored
by its integrability \cite{Katana89B}

The developed geometric construction in the theory of defects can be inverted,
and we can consider the gravity interaction of masses in the Universe as
the interaction of defects in elastic ether. Then point masses and cosmic
strings \cite{VilShe94} correspond to point defects (vacancies and
impurities) and wedge dislocations. We have the question in this framework
about the elastic gauge which has direct physical meaning in the geometric
theory of defects. If we take the standpoint of the theory of defects then
the elastic properties of ether correspond to some value of the Poisson
ratio which can be measured experimentally.

It seems interesting and important for applications to include time in the
considered static approach for describing motion of defects in the media.
Such a model is absent at present. From geometric point of view this
generalization can be easily performed at least in principle. It is
sufficient to change the Euclidean space $\MR^3$ to the Minkowski space
$\MR^{1,3}$ and to write a suitable Lagrangian quadratic in curvature and
torsion which corresponds to the true gravity model with torsion. One of the
arising problems is the physical interpretation of additional components of
the vielbein and Lorentz connection which contain the time index. The
physical meaning of the time component of the vielbein
$e_0{}^i\rightarrow\pl_0u^i=v^i$ is simple -- this is the velocity of a point
of media. This interpretation is natural from physical point of view because
motion of continuously distributed dislocations means flowing of media.
In fact, the liquid can be imagined as the elastic media with continuous
distribution of moving dislocations. It means that the dynamical theory of
defects based on Riemann--Cartan geometry must include hydrodynamics.
It is not clear at present how it could be. Physical interpretation of
the other components of the vielbein and the Lorentz connections with the
time index remains also obscured.

The author is very grateful to I.~V.~Volovich for numerous discussions.

\end{document}